\documentclass[cameraready]{Interspeech}  

\title{Time-Layer Adaptive Alignment for Speaker Similarity in Flow-Matching Based Zero-Shot TTS}

\author[affiliation={1}, equalcontribution]{Haoyu}{Li}
\author[affiliation={2}, equalcontribution]{Mingyang}{Han}
\author[affiliation={1}]{Yu}{Xi}
\author[affiliation={2}]{Dongxiao}{Wang}
\author[affiliation={1}]{Hankun}{Wang}
\author[affiliation={2}]{Haoxiang}{Shi}
\author[affiliation={2}]{Boyu}{Li}
\author[affiliation={2}]{Jun}{Song}
\author[affiliation={2}]{Zheng}{Bo}
\author[affiliation={3}, correspondingauthor]{Shuai}{Wang}
\author[affiliation={1}, correspondingauthor]{Kai}{Yu}

\address{
    $^1$ X-LANCE Lab, MoE Kay Lab of Artificial Intelligence, School of Computer Science,\\Shanghai Jiao Tong University, Shanghai, China\\
    $^2$ Taobao \& Tmall Group of Alibaba, China\\
    $^3$ School of Intelligence Science and Technology, Nanjing University, Suzhou, China
}

\email{\{haoyu.li.cs,kai.yu\}@sjtu.edu.cn, hanmingyang.hmy@alibaba-inc.com, shuaiwang@nju.edu.cn}


\usepackage{comment}
\usepackage{amsmath,graphicx,hyperref}
\usepackage{booktabs}
\usepackage{pifont}
\usepackage{multicol,multirow}
\usepackage{amsfonts}
\usepackage{hyperref}
\usepackage{subcaption}

\usepackage{url,hyperref,cleveref}
\begin{document}

\maketitle

\begin{abstract}
    Flow-Matching (FM)-based zero-shot text-to-speech (TTS) systems exhibit high-quality speech synthesis and robust generalization capabilities. However, the speaker representation ability of such systems remains underexplored, primarily due to the lack of explicit speaker-specific supervision in the FM framework. To this end, we conduct an empirical analysis of speaker information distribution and reveal its non-uniform allocation across time steps and network layers, underscoring the need for adaptive speaker alignment. Accordingly, we propose \textit{Time-Layer Adaptive Speaker Alignment} (TLA-SA), a strategy that enhances speaker consistency by jointly leveraging temporal and hierarchical variations. Experimental results show that TLA-SA substantially improves speaker similarity over baseline systems on both research- and industrial-scale datasets and generalizes well across diverse model architectures, including decoder-only language model (LM)-based and free TTS systems. A demo is provided\footnote{https://tlasa-tts.github.io/}.
\end{abstract}

\section{Introduction}
Driven by advances in large-scale datasets and model scaling, modern Text-to-Speech (TTS) systems have achieved remarkable zero-shot generation, enabling them to synthesize speech for unseen speakers without fine-tuning. These models excel at capturing speaker characteristics, such as timbre and prosody, thereby enhancing their utility in real-world scenarios.

Recent state-of-the-art (SOTA) TTS frameworks~\cite{arxiv2024-anastassiou-seedtts,arxiv2024-duzhihao-cosyvoice,arxiv2025-duzhihao-cosyvoice2,arxiv2025-duzhihao-cosyvoice3,arxiv2025-zhousiyi-indextts2,arxiv2025-zhangbowen-minimax,acl2025-chenyushen-f5tts} increasingly leverage Flow Matching (FM), owing to its superior distributional modeling and robust generalization.
Existing FM-based TTS architectures generally follow two paradigms:
(1) conditioned on upstream intermediate representations (e.g., LM-generated tokens) to learn token-to-spectrogram mappings;
(2) conditioned only on text (with optional prompts) to directly generate spectrograms.
The first paradigm employs an FM decoder conditioned on speech-text LM outputs~\cite{arxiv2024-anastassiou-seedtts,arxiv2024-duzhihao-cosyvoice,arxiv2025-duzhihao-cosyvoice2,arxiv2025-duzhihao-cosyvoice3,arxiv2025-zhousiyi-indextts2,arxiv2025-zhangbowen-minimax}.
Specifically, Seed-TTS~\cite{arxiv2024-anastassiou-seedtts} leverages large-scale autoregressive and diffusion-based architectures to achieve human-level naturalness and robust zero-shot generalization. CosyVoice series~\cite{arxiv2024-duzhihao-cosyvoice,arxiv2025-duzhihao-cosyvoice2,arxiv2025-duzhihao-cosyvoice3} achieve high naturalness and low latency, while IndexTTS2~\cite{arxiv2025-zhousiyi-indextts2} enhances prosodic control via duration and emotion disentanglement. Minimax~\cite{arxiv2025-zhangbowen-minimax} further utilizes a learnable speaker encoder and auxiliary FM to strengthen zero-shot generalization. 
In the second type, the FM TTS model generates a Mel spectrogram conditioned on text and speech~\cite{acl2025-chenyushen-f5tts,icassp2024-yiweiguo-voiceflow,slt2024-eskimez-e2tts} without an LM. VoiceFlow~\cite{icassp2024-yiweiguo-voiceflow} introduces a rectified FM algorithm that enhances generation efficiency.
E2-TTS~\cite{slt2024-eskimez-e2tts} facilitates end-to-end FM-based synthesis via blank token padding, while F5-TTS~\cite{acl2025-chenyushen-f5tts} enhances this approach with a dedicated text encoder and optimized sampling to achieve high real-time efficiency.

Despite the success of FM-based frameworks, the standard training objective remains limited, as it implicitly models the speech distribution without explicitly enforcing perceptual attributes like speaker identity.
This leads to sub-optimal fine-grained feature refinement, particularly speaker similarity, in zero-shot scenarios. Consequently, auxiliary supervision is required to provide explicit guidance for these attributes.
Prior works~\cite{iclr2025-yusihyun-repa,iccv2025-tri-taro,is2025-jeongsoo-f5tts_ctc_speaker_areg,arxiv2026-wu-archi} show that external representation alignment applied to intermediate FM features accelerates convergence and improves performance. In the context of TTS, previous work aligns text and speech jointly~\cite{is2025-jeongsoo-f5tts_ctc_speaker_areg} or in text-only modalities~\cite{arxiv2026-wu-archi} to increase efficiency.
However, two issues remain underexplored: (1) the distribution of speaker information across denoising trajectories and hierarchical layers, and (2) the optimal integration of speaker supervision to leverage these dynamics.
To bridge the gap, we first characterize the speaker information flow within FM-based architectures, revealing significant variance across both layers and denoising timesteps. These findings motivate the need for adaptive alignment. We thus propose \textit{Time-Layer Adaptive Speaker Alignment} (TLA-SA), a mechanism that dynamically aligns latent FM representations with speaker embeddings from a pre-trained encoder, conditioned on both the denoising timestep and architectural depth.
Our contributions are threefold:
\begin{itemize}
\item We propose the TLA-SA training strategy to enhance speaker consistency in zero-shot TTS. By integrating temporal and hierarchical information, TLA-SA provides dynamic supervision, effectively aligning intermediate FM representations with a pre-trained speaker latent space.
\item We conduct an analytical study demonstrating that speaker-specific information is non-uniformly distributed across denoising steps and model layers. This highlights the necessity of adaptive alignment and provides critical insights into the internal dynamics of speaker modeling in FM-based TTS.
\item Extensive experimental evaluations confirm that the proposed TLA-SA consistently improves speaker fidelity across varying data scales (small- to large-scale) and diverse mainstream model architectures, demonstrating its generalization.
\end{itemize}

\begin{figure}[t]
    \centering
    \includegraphics[width=1.0\linewidth]{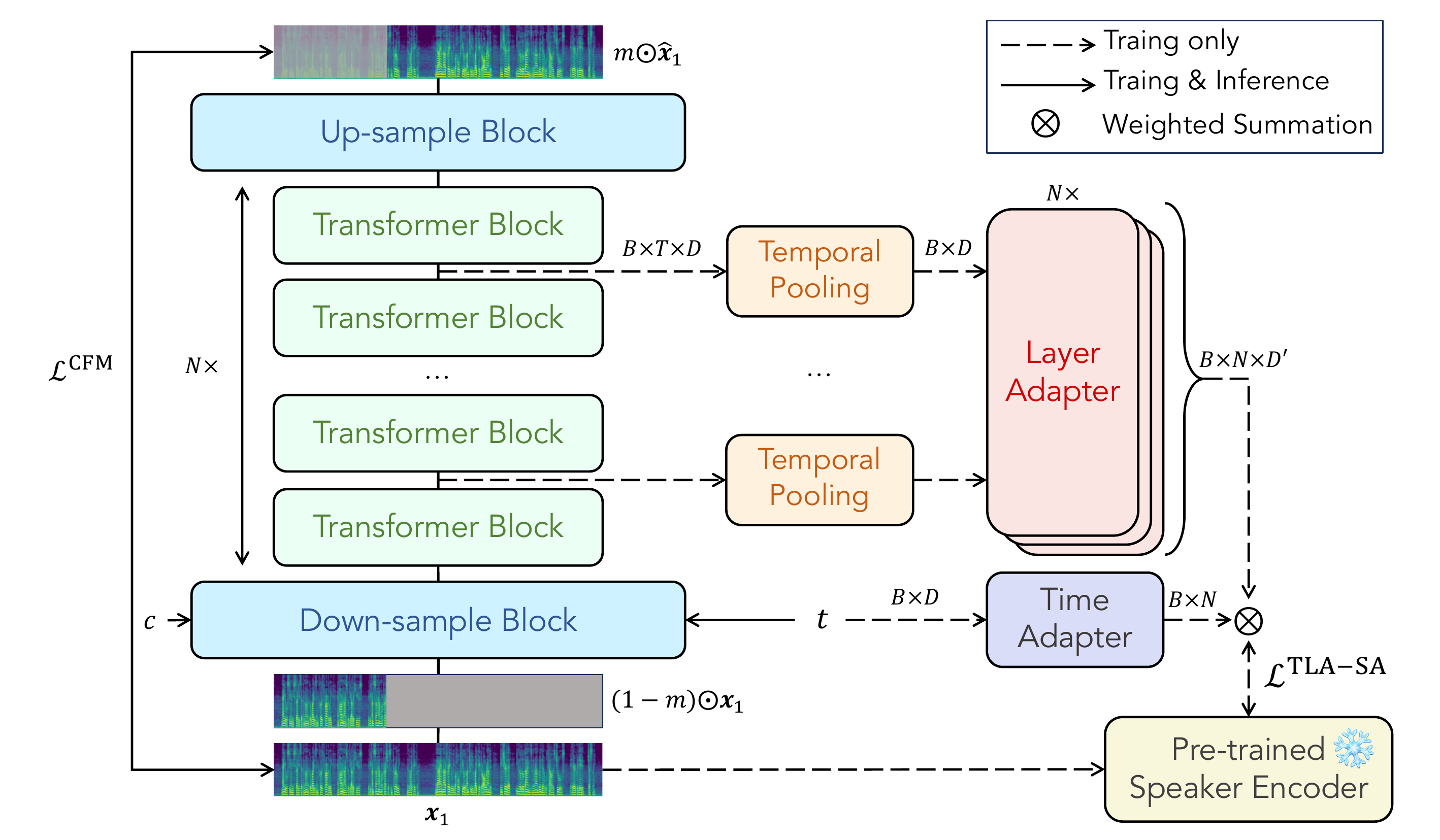}
    \caption{Overview of TLA-SA. A U-Net-based FM model is illustrated, and similar integration patterns apply to other architectures. 
    TLA-SA enforces speaker alignment across intermediate Transformer blocks, incorporating supervision on both timestep and model layer dimensions. }
    \vspace{-15pt}
    \label{fig:overview}
\end{figure}

\section{Methodology}
\label{sec:methodology}
\subsection{Flow-Matching based zero-shot TTS}
Flow-Matching (FM) zero-shot TTS models reconstruct Mel spectrograms through a mask prediction strategy, as shown in~\Cref{fig:overview}. During training, the FM model learns a flow $\phi_t(\mathbf{x}_t, c)$ that transforms a noise distribution $\mathbf{x}_0 \sim \mathcal{N}(\mathbf{0}, \mathbf{I})$ into the data distribution $\mathbf{x}_1 \sim q(\mathbf{x})$, conditioned on the masked input $\mathbf{x}_t$ and speaker embedding $c$. Given a vector field parameterized by $\theta$, the Conditional Flow Matching (CFM) loss minimizes the $\ell_2$ distance along the optimal transport path:
\begin{align}
    \mathcal{L}^{\text{CFM}}(\theta) = \mathbb{E}\left\Vert\mathbf{m} \odot [\mathbf{v}_t(\phi_t({\mathbf{x}_t,c});\theta) - (\mathbf{x}_1 - \mathbf{x}_0) ]\right\Vert_2^2,
\end{align}
where $t \sim \mathcal{U}(0,1)$ is the denoising time, and $\mathbf{m}$ denotes the mask.
During inference, Gaussian noise $\mathbf{x}_0$ is sampled, and an ODE conditioned on text input $\mathbf{x}$ and speaker condition $\mathbf{c}$ is solved to generate a new sample:
\begin{align}
    \hat{\mathbf{x}}_1=\mathbf{{m}} \odot \int_{0}^1 \mathbf{v}_t(\phi_t(\text{Concat}(\mathbf{x},\mathbf{x}_0),\mathbf{c});\theta) dt.
    \label{equ:fm-ode}
\end{align}
This formulation enables zero-shot generalization, allowing the model to synthesize speech for speakers specified by $c$, even if they are unseen during training.

\subsection{Speaker distribution in the FM-based TTS}
\label{sec:cknna-analysis}
\begin{figure}[ht]
  \centering
  \includegraphics[width=0.6\linewidth]{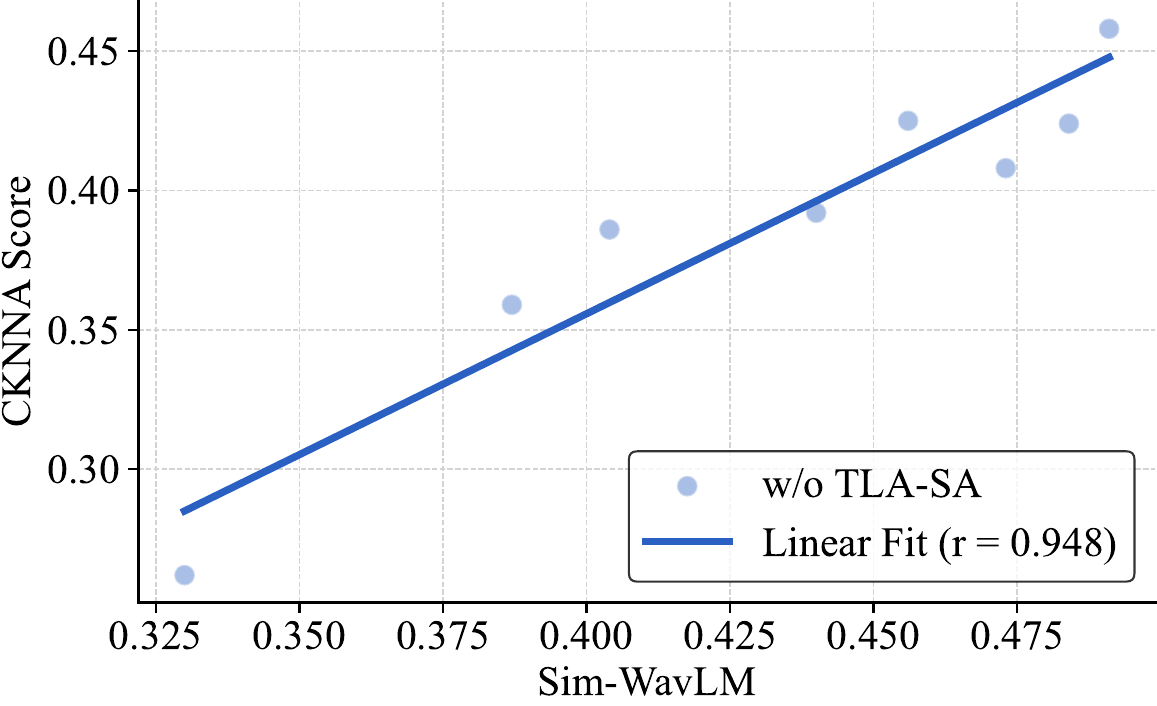}
  \caption{Correlation between speaker similarity and CKNNA. Each point denotes a checkpoint at a distinct training step.}
  \label{fig:cknna-spksim}
\end{figure}

To analyze the distribution of speaker information in the FM-based TTS model, we adopt the training-free Centered Kernel Nearest-Neighbor Alignment (CKNNA)~\cite{arxiv-huh-platonic} criterion, which quantifies the similarity between two sets of embeddings.
We first examine whether CKNNA faithfully reflects the speaker modeling quality of TTS models.
To evaluate baseline checkpoints across various training steps, we first compute CKNNA speaker similarity for $B$ synthesized utterances using a pre-trained speaker encoder.
Specifically, let $E_i \in \mathbb{R}^{B \times T \times D}$ denote the intermediate representation from the $i$-th layer of an $N$-layer FM module. The corresponding speaker representation is derived via temporal mean pooling as follows:
\begin{align}
    E^{\text{Speaker}}_i=\text{AvgPooling}_{T}(E_i) \in \mathbb{R}^{B\times D}.
    \label{equ:fm-pred-speaker-embedding}
\end{align}
Reference speaker embeddings $E^{\text{SA}} \in \mathbb{R}^{B \times D^{\prime}}$ are extracted from $B$ speech prompts, which comprise 37 distinct speakers and 500 utterances, using the same pre-trained encoder.
To quantify the speaker information captured at the $i$-th layer, we compute the CKNNA score between $E_i^{\text{Speaker}}$ and $E^{\text{SA}}$, reflecting the alignment of speaker features.
As shown in Fig.~\ref{fig:cknna-spksim}, CKNNA exhibits a strong positive linear correlation with speaker similarity, validating its utility as a metric for analyzing speaker information.

\begin{figure}[hbt]
  \centering
  \begin{subfigure}[b]{0.22\textwidth}
    \centering
    \includegraphics[width=\textwidth]{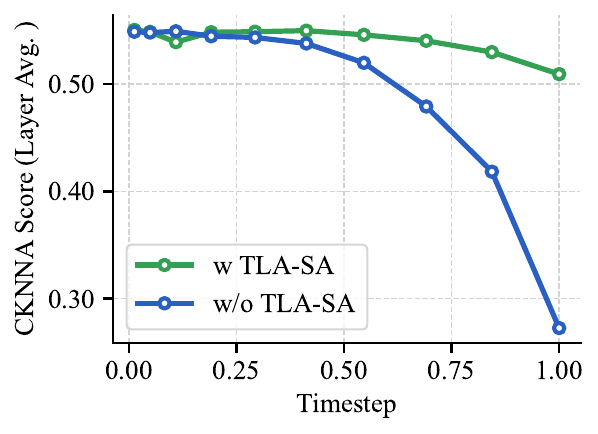}
    \caption{Timestep -- CKNNA Score.}
  \end{subfigure}
  \begin{subfigure}[b]{0.22\textwidth}
    \centering
    \includegraphics[width=\textwidth]{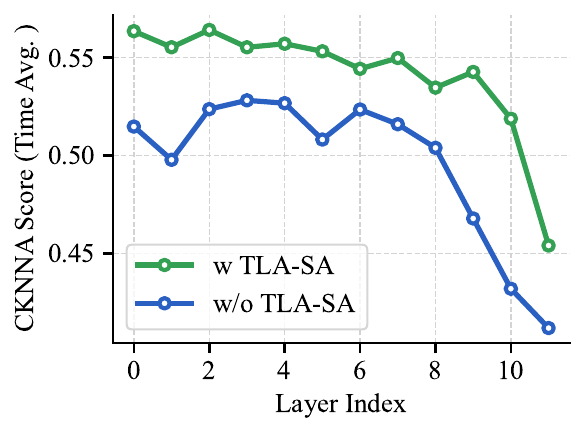}
    \caption{Layer Index -- CKNNA Score.}
  \end{subfigure}
  \caption{ CKNNA varies across time and layers on the converged checkpoint. }
  \label{fig:cknna-layer-timestep}
\end{figure}

Leveraging this metric, we analyze the speaker information distribution within the baseline FM model via CKNNA, revealing significant non-uniformity across both denoising timesteps and model layers, as illustrated in~\Cref{fig:cknna-layer-timestep}.
Along the temporal dimension, we observe that speaker information is predominantly encoded during the initial denoising steps, where noise levels are highest. This observation is consistent with the understanding that the early stages of the FM prioritize establishing global characteristics like speaker identity.
As for the hierarchical dimension, outputs of layers differ in their capacity to represent speaker information, with the final layers being particularly weak.
Results indicate that a fixed and uniform supervision strategy is suboptimal, as it fails to adapt to these dynamics. This motivates an adaptive supervision scheme that accounts for both the timestep and layer when enhancing speaker modeling.

\begin{table*}[th]
    \centering
    \caption{Performance comparison on Seed-TTS test sets. \textbf{\#Tr. Data}: total hours of training data. \textbf{SFT}: fine-tuning stage on speaker-related, task-specific data. \textbf{Bold results} indicate the best performance of our models.}
    \begin{resizebox}{1.0\linewidth}{!}{
    \begin{tabular}{lcc|ccc|ccc}
        \toprule
        \multirow{2}{*}{\textbf{Models}} & \multirow{2}{*}{\textbf{\#Tr. Data}} & \multirow{2}{*}{\textbf{SFT}} & \multicolumn{3}{c|}{\textbf{\textit{test-zh}}} & \multicolumn{3}{c}{\textbf{\textit{test-en}}} \\
        \cmidrule{4-9}
         & & & \textbf{CER (\%) $\downarrow$} & \textbf{Sim-WavLM $\uparrow$} & \textbf{Sim-ERes2Net $\uparrow$} & \textbf{WER (\%) $\downarrow$} & \textbf{Sim-WavLM $\uparrow$} & \textbf{Sim-ERes2Net $\uparrow$} \\
        \midrule
            \textbf{Human} & - & - & 1.26 & 0.755 & 0.775 & 2.14 & 0.734 & 0.742 \\
        \midrule
            \textbf{SparkTTS}~\cite{arxiv2025-wangxinsheng-sparktts} & 100k & \ding{56} & 1.20 & 0.672 & - & 1.98 & 0.584 & - \\
            \textbf{FireRedTTS}~\cite{arxiv2024-guohanhan-fireredtts} & 150k & \ding{51} & 1.51 & 0.635 & 0.653 & 3.82 & 0.460 & 0.526 \\
            \textbf{FireRedTTS-2}~\cite{arxiv2025-xie-fireredtts2} & 500k & \ding{51} & 1.14 & 0.736 & - & 1.95 & 0.665 & - \\ 
            \textbf{CosyVoice 2}~\cite{arxiv2025-duzhihao-cosyvoice2} & 200k & \ding{51} & 1.45 & 0.748 & 0.806 & 2.57 & 0.652 & 0.736 \\
        \midrule
            \textbf{CosyVoice 2} & 100k & \ding{56} & \textbf{1.31} & 0.714 & 0.761 & \textbf{2.48} & 0.606 & 0.677 \\
            \textbf{CosyVoice 2 + TLA-SA} & 100k & \ding{56} & 1.34 & \textbf{0.736} & \textbf{0.775} & 2.57 & \textbf{0.644} & \textbf{0.705} \\
        \bottomrule
    \end{tabular}
    }\end{resizebox}
    \label{tab:large_scale}
\end{table*}

\vspace{-10pt}
\subsection{Time and layer adaptive speaker alignment}
\label{sec:method-tlasa}
We introduce \textit{Time-Layer Adaptive Speaker Alignment} (TLA-SA), which facilitates both temporal and hierarchical information to boost speaker modeling in FM-based TTS.
The core mechanism involves dynamically aligning intermediate FM representations with speaker embeddings via a pre-trained encoder, thereby enforcing consistency in speaker identity.
The speaker alignment loss for layer i, denoted by $\mathcal{L}^{\text{SA}}_i$, minimizes the discrepancy between the predicted embedding $E^{\text{Speaker}}_i$ (defined in~\Cref{equ:fm-pred-speaker-embedding}) and the reference speaker embedding $E^{\text{SA}}$ extracted from a pre-trained encoder:
\begin{align}
    \mathcal{L}^{\text{SA}}_{i} = \mathcal{D}(E^{\text{SA}}, \text{MLP}^{\text{Layer}}_i(E^{\text{Speaker}}_{i})),
\end{align}
where $\mathcal{D}(\cdot,\cdot)$ is the cosine similarity function and $\text{MLP}^{\text{Layer}}_i$ denotes one of $N$ layer-wise adapters.
To incorporate temporal dynamics into the hierarchical speaker alignment, we utilize the denoising timestep to predict a set of adaptive weights $\mathbf{w} \in \mathbb{R}^{B\times N}$. These weights are used to aggregate layer-wise SA losses across $N$ blocks. Specifically, at timestep $t$, the time embedding $\text{Emb}^{\text{Time}}$ is processed by a time adapter $\text{MLP}^{\text{Time}}$, which computes the weights via a softmax operation across the layer dimension:
\begin{align}
    \mathbf{w} = \text{Softmax}_{N}( \text{MLP}^{\text{Time}}(\text{Emb}^{\text{Time}}(t))).
\end{align}
Accordingly, TLA-SA loss can be formulated as follows:
\begin{align}
    \mathcal{L}^{\text{TLA-SA}} &= \sum_{i=0}^{N-1}\mathbf{w}_{i}\mathcal{L}^{\text{SA}}_{i}  + \alpha\mathcal{L}^{\text{Reg}} (\mathbf{w}), \\
    \mathcal{L}^{\text{Reg}} &= - \text{Entropy}(\mathbf{{w}}).
\end{align}
We add an entropy penalty regularization term $\mathcal{L}^{\text{Reg}}$~\cite{neurips2020-zhaoshanshan-entropy_reg} to stabilize weight learning and set $\alpha=0.01$ empirically. The overall training objective is given by:
\begin{align}
    \mathcal{L} &= \mathcal{L}^{\text{CFM}} + \lambda\mathcal{L}^{\text{TLA-SA}},
\end{align}
with $\lambda=0.5$ controlling the trade-off between the two terms.

\section{Experimental Setups}
\subsection{Datasets}
To evaluate the proposed method across varying data scales, we conduct experiments on the following datasets:
\begin{itemize}
    \item \textbf{TLASA-100k}: An industrial corpus comprising 100k hours of speech (60k Mandarin, 40k English), primarily sourced from Emilia~\cite{emilia}, WenetSpeech~\cite{wenetspeech}, and WenetSpeech4TTS~\cite{wenetspeech4tts}. The final set was derived via speaker clustering and de-duplication. The data list will be available\footnote{https://github.com/TLASA-TTS/TLASA-100k}.

    \item \textbf{Seed-TTS}~\cite{arxiv2024-anastassiou-seedtts}: A standard benchmark for zero-shot TTS evaluation in Mandarin and English. For models trained on large-scale data, we utilize the \textit{test-zh} and \textit{test-en} subsets, which consist of 2,020 and 1,088 prompt-text pairs.

    \item \textbf{LibriTTS}~\cite{libritts}: A high-quality English corpus (585 hours) used for small-scale assessment. Following \cite{arxiv2025-duzhihao-cosyvoice2, aaai24-duchenpeng-unicats}, we evaluate 500 utterances from the \textit{test-clean} split across 37 speakers. This set also facilitates our CKNNA-based analysis.    
\end{itemize}

\subsection{Model structures and training configurations}

\textbf{Model structures.} To assess the effectiveness and structural generalization of TLA-SA, we conduct experiments on two distinct architectures:
\textbf{LM-based FM TTS:} Following CosyVoice~2~\cite{arxiv2025-duzhihao-cosyvoice2}, our model integrates a speech-text LM with a Transformer U-Net CFM decoder (2 down-sampling, 12 middle, and 2 up-sampling layers). TLA-SA is applied to the middle layers.
\textbf{LM-free FM TTS.} We employ an FM-based framework with a transformer text encoder and 9 MMDiT~\cite{icml2024-esser-mmdit} decoder layers, similar to~\cite{acl2025-chenyushen-f5tts}.

\textbf{Training Configuration.} All models are trained from scratch using 24kHz speech. We employ the AdamW~\cite{arxiv2017-loshchilov-adamw} optimizer ($\beta_1=0.9, \beta_2=0.999$) with a cosine learning rate scheduler and a 0.25-epoch warm-up. For the \textbf{TLASA-100k} set, we train for 156k steps with a total batch size of 64 (batch 16 and 4 gradient accumulations), decaying the learning rate from 8e-4 to 1e-6. For \textbf{LibriTTS}, models are trained for 125k steps with a batch size of 16 and a learning rate range of 2e-4 to 1e-4. The TLA-SA loss utilizes a WavLM-based encoder fine-tuned for speaker verification~\cite{taslp2022-chensanyuan-wavlm_sv}.

\subsection{Evaluation Metrics}
\textbf{WER/CER:} We report the Word Error Rate (WER) for English and Character Error Rate (CER) for Mandarin using the pre-trained ASR models provided by Seed-TTS~\cite{arxiv2024-anastassiou-seedtts}. 
\textbf{Speaker Similarity:} To assess speaker consistency, we compute the cosine similarity between the embeddings of the generated speech and the original prompt. We utilize two robust pre-trained models: a speaker-refined WavLM~\cite{arxiv2024-anastassiou-seedtts} (\textbf{Sim-WavLM}) and the ERes2Net~\cite{icassp2025-chenyafeng-3dspeaker} model as configured in CosyVoice~2~\cite{arxiv2025-duzhihao-cosyvoice2} (\textbf{Sim-ERes2Net}).
\textbf{Mean Opinion Score (MOS):} Twelve speech experts conducted a double-blind evaluation of 20 utterance pairs (10 Mandarin, 10 English) from Seed-TTS. To ensure unbiased assessment, listeners rated speech against prompts on a 1--5 scale, with higher scores indicating superior perceptual similarity to the target speaker.

\section{Results and Analysis}
\label{sec:results}
\subsection{TLA-SA on large-scale dataset}

Results in~\Cref{tab:large_scale} show that TLA-SA yields average absolute gains of 3.0\% in Sim-WavLM and 2.1\% in Sim-ERes2Net over our baseline with comparable CER/WER. These results, aligning with~\Cref{fig:cknna-layer-timestep}, demonstrate that TLA-SA provides a theoretically-grounded approach to improving speaker modeling directly within the pre-training phase.
Compared to SparkTTS~\cite{arxiv2025-wangxinsheng-sparktts} at equivalent data scales, our TLA-SA enhanced model achieves superior performance.
Notably, it significantly outperforms FireRedTTS~\cite{arxiv2024-guohanhan-fireredtts} despite the latter using 50\% more data and a dedicated speaker SFT stage. This highlights the ability of TLA-SA to improve speaker identity preservation without requiring complex supervised fine-tuning. Furthermore, our model matches the speaker similarity of FireRedTTS-2~\cite{arxiv2025-xie-fireredtts2}, which utilizes $4\times$ the data and SFT.
Finally, our model with TLA-SA achieves comparable performance with the official 150k-hour CosyVoice~2~\cite{arxiv2025-duzhihao-cosyvoice2}, maintaining only a 2.1\% gap on average against the model using $2\times$ data and SFT.
These results establish TLA-SA as a data-efficient, SFT-free approach to enhancing speaker modeling.
\vspace{-8pt}
\begin{figure}[h]
    \centering
    \includegraphics[width=0.65\linewidth]{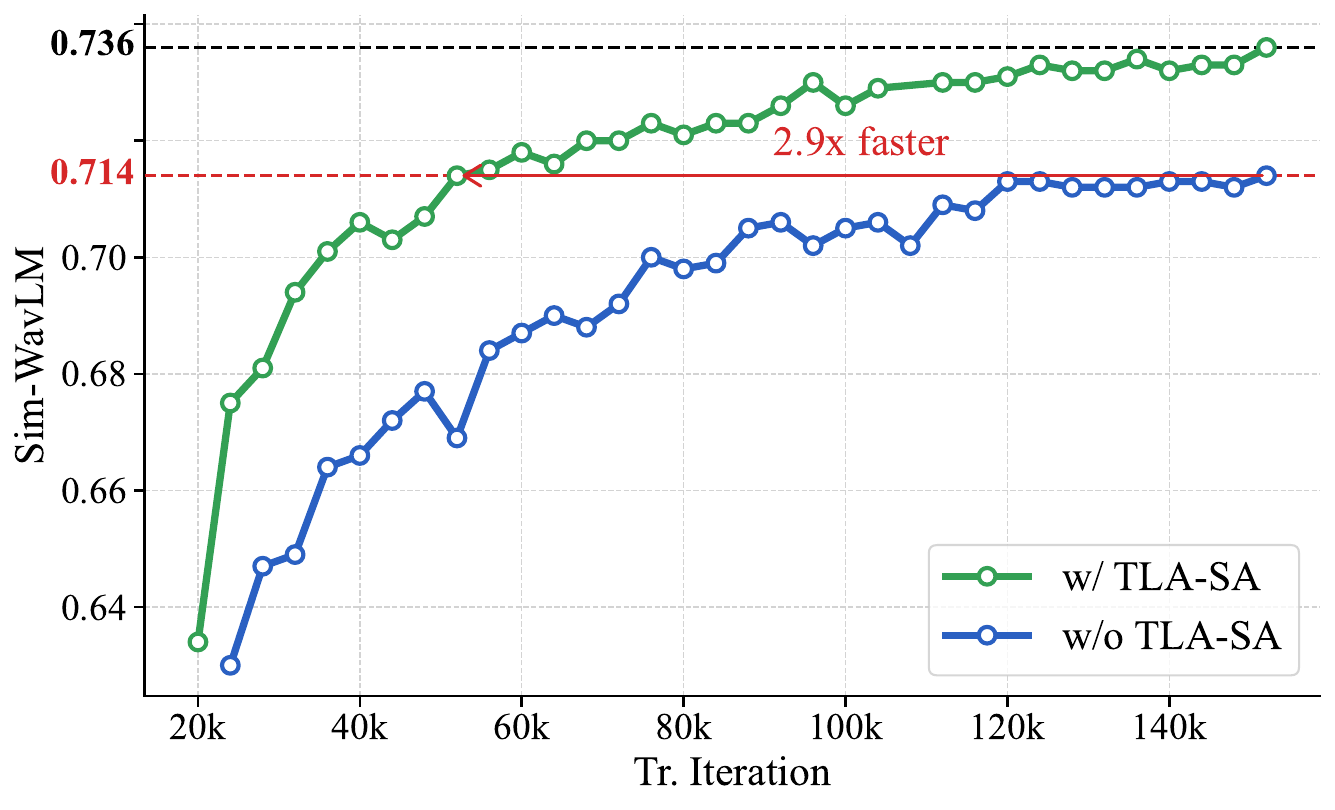}
    \vspace{-5pt}
    \caption{Sim-WavLM across steps on Seed-TTS \textit{test-zh}.}
    \label{fig:convergence}
\end{figure}
\vspace{-7pt}

Besides, the benefits of TLA-SA also manifest in the optimization process. As illustrated in \Cref{fig:convergence}, TLA-SA enhances speaker modeling with a 2.2\% Sim-WavLM gain and a 2.9$\times$ convergence speedup on TLASA-100k. This dual advantage demonstrates the efficacy of TLA-SA in providing structured and efficient supervision for speaker representation.

\vspace{-5pt}
\subsection{TLA-SA on high-quality LibriTTS}
\label{sec:small_scale}
We evaluate TLA-SA on the high-quality LibriTTS dataset, which is only 0.6\% the size of TLASA-100k. As shown in~\Cref{tab:small_scale}, TLA-SA consistently outperforms existing speaker alignment baselines across both Sim-WavLM and Sim-ERes2Net metrics. These performance gains demonstrate that jointly leveraging temporal and hierarchical dynamics is essential for effective speaker alignment in FM-based TTS. Notably, despite utilizing only a single speaker encoder for supervision, TLA-SA enhances global speaker modeling without overfitting to the specific supervisor's feature space. This cross-metric improvement underscores the robust generalization of our approach across different evaluation models.

\vspace{-5pt}
\begin{table}[h]
    \centering
    \caption{Comparison of TLA-SA against existing speaker alignment methods on LibriTTS.}
    \vspace{-5pt}
    \begin{resizebox}{1.0\linewidth}{!}{
    \begin{tabular}{l|cc|cc}
    \toprule
        \multicolumn{1}{l|}{\textbf{SA Method}} & \textbf{Time Adapt.} & \textbf{Layer Adapt.} & \textbf{Sim-WavLM $\uparrow$} & \textbf{Sim-ERes2Net $\uparrow$} \\
    \midrule
        None & \ding{56} & \ding{56} & 0.510 & 0.571 \\        
    \midrule
        TRA~\cite{iccv2025-tri-taro} & \ding{51} & \ding{56} & 0.513 & 0.573 \\
        Speech Ailgn~\cite{is2025-jeongsoo-f5tts_ctc_speaker_areg} & \ding{56} & \ding{51} & 0.524 & 0.583 \\
        TLA-SA & \ding{51} & \ding{51} & \textbf{0.538} & \textbf{0.596} \\
    \bottomrule
    \end{tabular}
    }\end{resizebox}
    \label{tab:small_scale}
\end{table}
\vspace{-10pt}

\begin{figure}[bt]
  \centering
  \begin{resizebox}{0.82\linewidth}{!}{
  \begin{subfigure}[b]{0.22\textwidth}
    \centering
    \includegraphics[width=\textwidth]{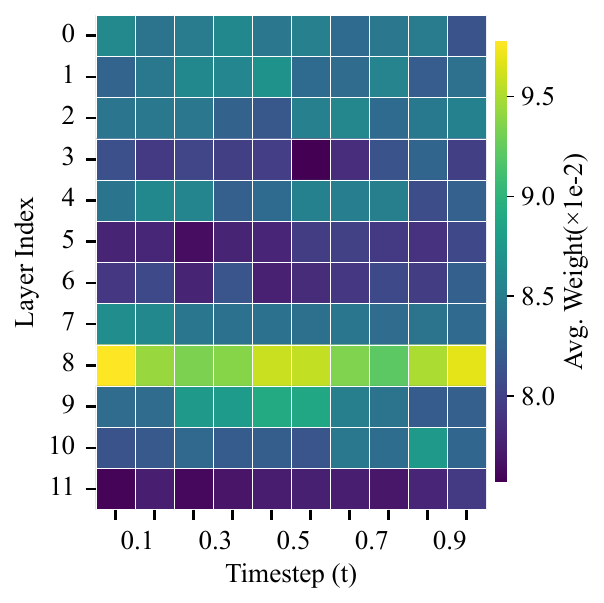}
    \caption{Random initialization.}
  \end{subfigure}
  \begin{subfigure}[b]{0.22\textwidth}
    \centering
    \includegraphics[width=\textwidth]{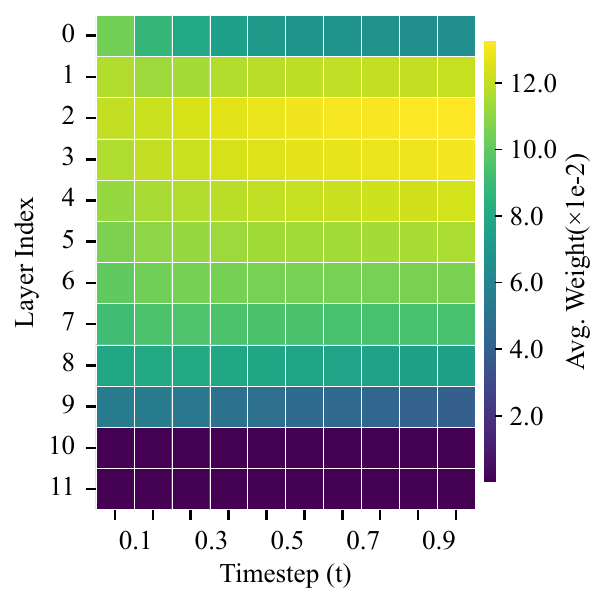}
    \caption{Learned weights.}
  \end{subfigure}
  }\end{resizebox}
  \vspace{-5pt}
  \caption{Heatmap of adaptive $\mathbf{w}$ at different training stage.}
  \label{fig:tla-w}
\end{figure}

Furthermore, we visualize the adaptive weights $\mathbf{w}$ across both denoising timesteps and model layers. As illustrated in \Cref{fig:tla-w}, TLA-SA applies non-uniform supervision strength across the temporal and hierarchical dimensions, effectively integrating multi-scale information. The uneven weight distribution underscores the necessity of adaptive alignment, corroborating our observations in \Cref{fig:cknna-layer-timestep}. Specifically, higher weights are concentrated in shallower layers and later denoising stages, suggesting that these regions are more critical for speaker identity and thus require intensified supervision.

\subsection{Performance on different model structures}

To evaluate the structural generalization of TLA-SA, we conduct experiments on both LM-based (e.g., CosyVoice~\cite{arxiv2025-duzhihao-cosyvoice2}) and LM-free (e.g., F5-TTS~\cite{acl2025-chenyushen-f5tts}) FM-based TTS decoders. As shown in \Cref{tab:generalization}, TLA-SA consistently improves speaker modeling in both settings, demonstrating its broad architectural compatibility. Notably, on the LM-free architecture, TLA-SA yields substantial absolute gains in Sim-WavLM (0.458 vs. 0.398) and Sim-ERes2Net (0.571 vs. 0.500), confirming its ability to effectively guide speaker representation learning in the FM.

\vspace{-5pt}
\begin{table}[th]
    \centering
    \caption{Generalization of TLA-SA across model architectures on LibriTTS.}
    \begin{resizebox}{1.0\linewidth}{!}{
    \begin{tabular}{cc|cc}
    \toprule
        \textbf{Training Paradigm} & \textbf{TLA-SA} & \textbf{Sim-WavLM $\uparrow$} & \textbf{Sim-ERes2Net $\uparrow$} \\
    \midrule
        \multirow{2}{*}{LM-based} & \ding{56} & 0.510 & 0.571 \\
        & \ding{51} & \textbf{0.538} & \textbf{0.596} \\
    \midrule
        \multirow{2}{*}{LM-free} & \ding{56} & 0.398 & 0.500 \\
        & \ding{51} & \textbf{0.458} & \textbf{0.571} \\
    \bottomrule
    \end{tabular}
    }\end{resizebox}
    \label{tab:generalization}
\end{table}
\vspace{-10pt}

\subsection{Subjective evaluation on speaker similarity}

When trained on the TLASA-100k corpus, the proposed TLA-SA consistently outperforms the baseline, increasing MOS from 4.16 to 4.23 on \textit{test-zh} and from 3.98 to 4.10 on \textit{test-en}. These results demonstrate its efficacy in enhancing perceptual quality across both Chinese and English subsets.

\begin{table}[hbt]
    \centering
    \caption{Speaker similarity MOS score on Seed-TTS.}
    \begin{resizebox}{0.8\linewidth}{!}{
    \begin{tabular}{l|cc|c}
    \toprule
        \multicolumn{1}{c|}{\multirow{2}{*}{\textbf{Models}}} & \multicolumn{2}{c|}{\textbf{Datasets}} & \multirow{2}{*}{\textbf{Avg.}}\\
        \cmidrule{2-3}
          & \textbf{\textit{test-zh}} & \textbf{\textit{test-en}} \\
    \midrule
        CosyVoice 2 & 4.16 & 3.98 & 4.07\\
        CosyVoice 2 + TLA-SA & \textbf{4.23} & \textbf{4.10} & \textbf{4.17}\\
    \bottomrule
    \end{tabular}
    }\end{resizebox}
    \label{tab:placeholder}
\end{table}

\section{Conclusion}
\label{sec:conclusion}

This work introduces TLA-SA, a training strategy utilizing temporal and hierarchical speaker alignment to enhance speaker consistency in zero-shot FM-based TTS. By performing joint alignment across denoising timesteps and hierarchical layers, TLA-SA significantly strengthens speaker modeling. Our method achieves comparable or superior speaker similarity without specialized fine-tuning, maintaining effectiveness across varying data scales. Furthermore, TLA-SA yields consistent gains for both LM-cascaded FM and standalone architectures, demonstrating robust architectural generalization. Future research will explore the applicability of TLA-SA to diverse FM variants and incorporate multi-modal alignment signals to further refine synthesis quality.

\section{Generative AI Use Disclosure}
We utilized generative AI tools exclusively for language editing and stylistic refinement (e.g., grammar, clarity, and concision). No AI systems contributed substantive technical content, experiment design, data analysis, figures, or results. All (co-)authors are responsible and accountable for the work and consent to submission. In accordance with ISCA policy, generative AI tools are not listed as co-authors and were not used to produce any significant portion of the manuscript.

\bibliographystyle{IEEEtran}
\bibliography{refs}

\end{document}